\documentclass[runningheads]{llncs}
\usepackage[T1]{fontenc}
\usepackage{graphicx}
\usepackage{amsmath}
\usepackage{amssymb}
\usepackage{booktabs}
\usepackage{multirow}
\usepackage{array}
\usepackage{algorithm}
\usepackage{algpseudocode}
\usepackage{xcolor}
\usepackage{url}
\usepackage{tikz}
\usetikzlibrary{arrows.meta,positioning,fit,calc,shapes.geometric,backgrounds}
\usepackage[colorlinks=true,linkcolor=blue,citecolor=blue,urlcolor=blue]{hyperref}

\begin{document}
\title{ExecuCritic: Calibrated Critic Shaping for\\
Code Generation with Verifiable Rewards}
\titlerunning{ExecuCritic: Calibrated Critic Shaping}
\author{Junjie Cao\inst{1} \and
Yingjie He\inst{2}}
\authorrunning{J. Cao et al.}
\institute{Intel Corporation, Beijing, China\\
\email{junjie.cao@intel.com}
\and
Peking University, Beijing, China}
\maketitle
\begin{abstract}
Execution feedback is a useful supervision signal for code models because
unit tests are objective and directly measure program correctness. Its
weakness is that an entire program is often reduced to one pass or fail
bit, leaving RLVR to solve a difficult credit assignment problem. At the
same time, coding systems often include separate reviewer or tester
roles, but these critics are usually prompted rather than trained and are
not calibrated against execution. We propose \emph{ExecuCritic}, a joint
training framework in which a coder and a critic are updated on the same
execution rollouts. The critic predicts pass or fail outcomes and gives
short diagnostic feedback; the coder uses this signal only when the
critic agrees with the executor on the current rollout group. Across
eight code benchmarks and two recent open backbones, ExecuCritic improves
over GRPO without a critic, prompted reviewer systems and scalar reward
model baselines, while requiring fewer policy gradient steps and fewer
sandbox executions. Ablations and reliability analyses suggest that the
gains come from better credit assignment rather than larger sampling
budgets.

\keywords{Code generation \and Verifiable reward \and Multi agent
learning \and Reinforcement learning \and Critic models.}
\end{abstract}
%
\section{Introduction}
\label{sec:intro}

Code language models~\cite{grattafiori2024llama} are increasingly evaluated as software agents rather
than as autocomplete systems~\cite{zhao2026masbench,zheng2026scilens}. In a realistic coding task, a model may
need to understand an issue, modify several files, run tests, read the
resulting trace and revise its patch. This workflow has made execution
feedback especially attractive: a unit test is cheap compared with human
annotation, and its result is directly tied to whether the program works.
Reinforcement learning with verifiable rewards (RLVR) builds on this
property by fine tuning code models with policy gradient methods such as
GRPO~\cite{shao2024deepseekmath} using rewards from sandboxed execution.
Recent code RL systems~\cite{guo2025deepseek,wei2026swe,%
zeng2025acecoder,liu2025code} show that even binary test feedback can
improve coding and reasoning ability, and parallel work on reward design
for LLM reasoning reports similar gains from denser or better aligned
signals~\cite{zheng2026gradients,shi2026spader,kang2026jpo}.

A second line of work improves code generation by decomposing the task
into roles. Several systems built around prompting~\cite{huang2023agentcoder,%
hong2024metagpt,islam2024mapcoder,madaan2023self} introduce
programmers, reviewers, testers or debuggers, so that a candidate program
can be checked before it is submitted. The analogy to ordinary software
development is clear: code is reviewed, tested and diagnosed rather than
written once and accepted. However, in most such systems the critic is
only a prompted role. It may produce fluent comments, but it is not
trained to match executor outcomes, and the coder is not trained to know
when those comments should be followed.

The two approaches therefore solve different parts of the problem. RLVR
has a trustworthy verifier, but the learning signal is thin. A long patch
receives the same binary credit on every generated token, even though the
failure may be caused by one branch condition, one API call or one
missing import. This leads to noisy updates whose effect is hard to
attribute to a specific mistake~\cite{zhu2026edis,jiang2026foe}, and
often encourages large rollout groups, which are expensive when every
candidate must be executed. Prompted systems with multiple roles provide richer language
feedback, but the feedback is not grounded in the actual executor. A
critic can be confident and wrong, and nothing in the training objective
teaches the coder to discount such advice. This motivates the question we
study in this paper: can execution rewards and learned critiques be
combined so that the critic makes RLVR less sparse without becoming an
untrusted reward model?

We propose \emph{ExecuCritic}, a joint training framework for a coder
agent~$\pi_c$ and a critic agent~$\pi_v$. The two agents share a frozen
backbone and use separate LoRA adapters~\cite{hu2022lora}, and both are
updated on the same pool of rollouts labelled by execution. For each
prompt, the coder samples candidate programs and a sandbox returns pass
or fail outcomes. The critic predicts these outcomes, ranks passing candidates
above failing ones and produces a short diagnosis tied to the observed
failure type. The coder is then trained with a calibrated advantage: the
standard GRPO signal is augmented with a critic score, but only to the
extent that the critic agrees with the executor within the current
rollout group. When the critic is unreliable, the update falls back
toward ordinary RLVR; when it becomes calibrated, its scores provide a
dense signal for credit assignment.

The critic is also not an extra frozen judge bolted on after training: it
has seen the same evolving distribution of code as the coder, so it can
rank candidates before execution and supply feedback when all of them
fail. In repository repair, where one sandbox call may install
dependencies and run many tests, this ranking cuts expensive executions
without simply enlarging the sampling budget.

We evaluate ExecuCritic with two recent open code backbones on benchmarks
spanning function synthesis, contamination controlled programming, library
use, competitive programming, repository issue resolution, cross-language
editing, patch ranking and test selection. Under matched data and token
budgets it improves over the strongest critic-free RLVR baseline by 3--4
absolute points and over prompted reviewer systems by 6--7, and reaches
that baseline's final accuracy with roughly $40\%$ fewer policy gradient
steps. Ablations tie the gains to the calibration gate, the pass/fail
margin and the diagnosis consistency objective rather than to extra
sampling.

\paragraph{Contributions.}
\begin{itemize}
    \item We formulate code agent training as joint optimisation of a coder
    and an execution-grounded critic updated on the same RLVR rollouts.
    \item We introduce a calibrated advantage that trusts critic scores only
    where they agree with executor outcomes, giving denser feedback with
    less risk of reward hacking.
    \item We train the critic to predict verdicts, rank passing above
    failing programs, and diagnose failures from executor traces.
    \item We study two backbones and eight benchmark views with ablations,
    seed checks, reliability, sample efficiency, budget and failure-type
    analyses.
\end{itemize}

\section{Related Work}
\label{sec:related}

\subsubsection{RL with verifiable rewards for code.} Following recent
work on RLVR for code~\cite{guo2025deepseek,wei2026swe,pan2024training,%
zeng2025acecoder,liu2025code,liu2026rstar}, a wave of studies has
applied executable feedback to code generation. These systems optimise
patching at repository scale or function synthesis with curated data
augmented by tests. They demonstrate the power of executable
supervision, but they still train a \emph{single} policy against a
\emph{single} scalar reward and therefore inherit the sparsity problem
studied in this paper. Related efforts outside code shape the reward
itself rather than the data, by aligning updates with gradient
evidence~\cite{zheng2026gradients}, by distributing group relative credit
across reasoning steps~\cite{shi2026spader}, or by replacing surface
overlap metrics with hierarchical task aware
rewards~\cite{wang2026ngrams,kang2026jpo}.

\subsubsection{Code generation with multiple agents.} A separate line of
work keeps the model frozen and orchestrates several roles through
prompting.
Self-Refine~\cite{madaan2023self} iterates between generation and
critique with one model; AgentCoder~\cite{huang2023agentcoder},
Meta-GPT~\cite{hong2024metagpt}, MapCoder~\cite{islam2024mapcoder} and
AutoGen-Coder~\cite{wu2024autogen} introduce dedicated programmer,
tester and reviewer agents. CodeT~\cite{chen2022codet} and
Reflexion~\cite{shinn2023reflexion} similarly use generated tests or
verbal feedback to filter or rewrite candidates. Learned coordination
policies decide which agent or action to invoke instead of fixing the
schedule in advance~\cite{jiang2026agentqmix,meng2026group}, and similar
role decompositions have been applied to multimodal decision
tasks~\cite{kang2026mmlegal}. These pipelines benefit
from role decomposition, but their critics are generally frozen and
uncalibrated. ExecuCritic instead trains the critic and coder jointly
under a verifiable reward.

\subsubsection{Critic and verifier models.} Process reward
models~\cite{lightman2024let} and
CriticGPT~\cite{mcaleese2024llm} demonstrate that critics trained
on labelled traces can rival much larger frozen judges. Outcome reward
models for math and code~\cite{cobbe2021training} extend
this idea to scalar correctness prediction. Analyses of how reasoning
models fail, whether by enumerating an error forest~\cite{jiang2026foe}
or by making the reasoning pathway explicit and
controllable~\cite{dong2026neureasoner}, suggest that a critic benefits
from naming the failure rather than only scoring it. Our work differs in
two key ways: the critic is trained \emph{simultaneously} with the policy whose
outputs it scores, and its score is fed back into the policy gradient
rather than used only during decoding.

\subsubsection{Reward shaping and dense feedback.} Reward shaping is a
classical idea in RL~\cite{ng1999policy}; recent LLM work explores
intrinsic rewards~\cite{yuan2024self}, learned dense
verifiers~\cite{wang2024math} and feedback generated by AI
systems~\cite{bai2022constitutional}. When several learning signals
disagree, reweighting them by their mutual consistency is often more
stable than summing them, an idea also used to resolve conflicting client
updates in distributed training~\cite{hong2026conflict}. ExecuCritic can
be viewed as reward shaping grounded by a verifier: the dense bonus is provided by a learned
critic, but the critic itself remains anchored to a verifiable executable
signal, which empirically helps prevent reward hacking.

\section{Method}
\label{sec:method}

\subsection{Problem setup}
\label{sec:method:setup}

We study a code generation setting in which a prompt~$x$ (a natural
language specification, a function signature, or a buggy file plus an
issue description) is mapped to a candidate solution~$y$ (a function body
or a unified diff). A black box executor $E(x,y)\in\{0,1\}$ returns~$1$ iff~$y$
passes all hidden unit tests for~$x$ and~$0$ otherwise. RLVR maximises
the expected execution reward
\begin{equation}
J(\pi)\;=\;\mathbb{E}_{x\sim\mathcal{D}}\,\mathbb{E}_{y\sim\pi(\cdot\mid x)}
\bigl[\,E(x,y)\,\bigr],
\label{eq:rlvr}
\end{equation}
typically with GRPO, which draws a group of $K$~rollouts
$\{y_i\}_{i=1}^{K}$ for each prompt and computes an advantage
$A_i=(R_i-\mu_R)/\sigma_R$ from rewards normalised within the group,
where $R_i=E(x,y_i)$. The advantage is then used in a clipped policy
gradient update~\cite{schulman2017proximal}.

ExecuCritic extends this setup with a second policy~$\pi_v$, the
\emph{critic}, which receives the prompt and a candidate solution and
produces both a critique in natural language~$c$ and a scalar
score~$\hat{s}\in[0,1]$. Both~$\pi_c$ and~$\pi_v$ are implemented as LoRA
adapters on top of a shared frozen backbone~$\pi_0$, so the additional
parameter cost over RLVR with only the coder remains modest. Unlike a
conventional reward model, $\pi_v$ is trained on the coder's evolving
output distribution, encouraging it to recognise the mistakes that the
current coder actually makes.

\subsection{Coder and critic architecture}
\label{sec:method:arch}

\begin{figure}[t]
\centering
\includegraphics[width=0.88\textwidth]{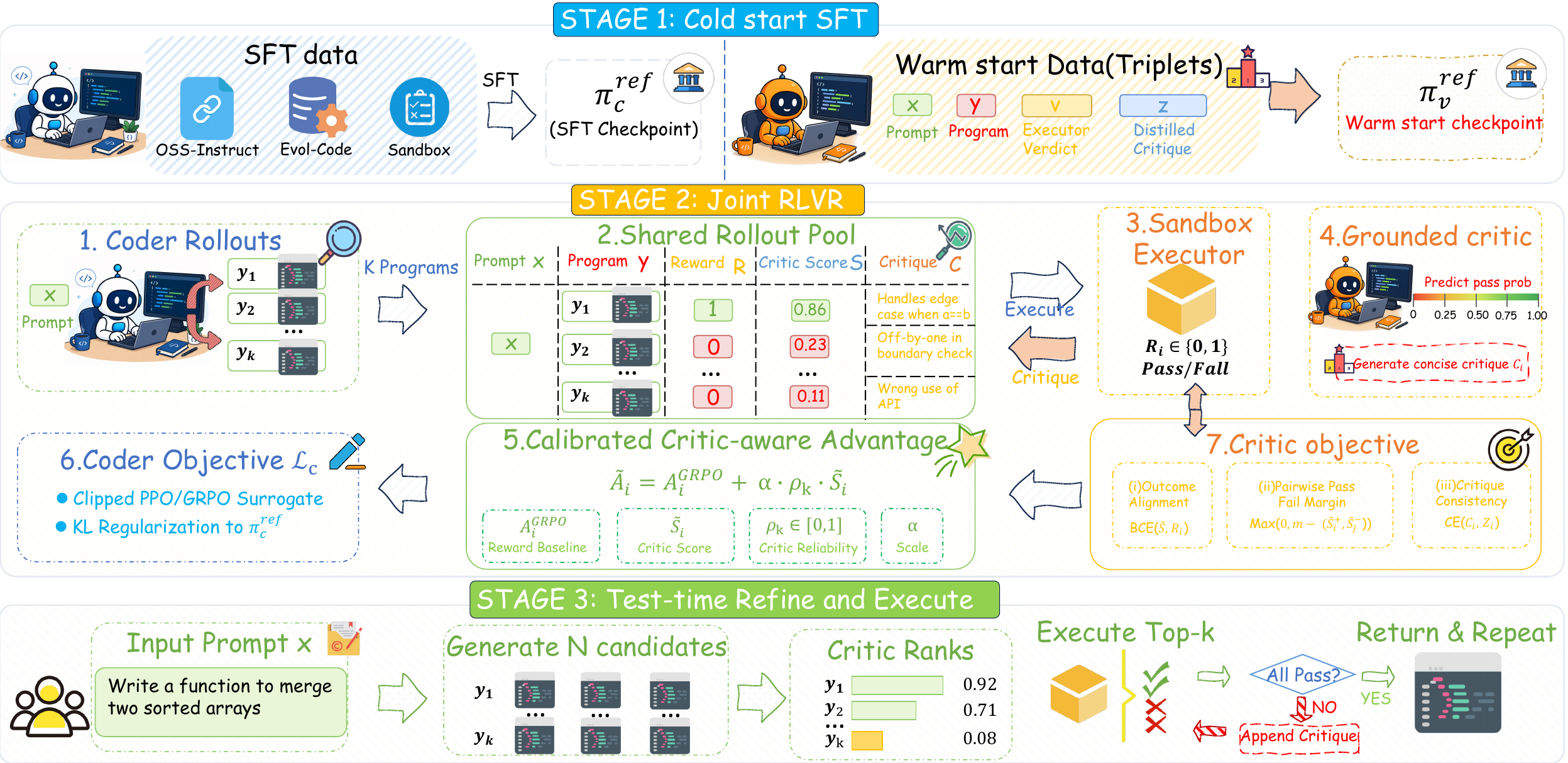}
\caption{Pipeline of \emph{ExecuCritic}. Stage 1 cold starts the coder with
SFT and the critic with warm start triplets of prompts, programs, executor
verdicts and distilled critiques. Stage 2 trains both on shared rollouts:
the sandbox gives pass/fail rewards, the critic predicts pass probability
and diagnoses failures, and the coder is updated with a calibrated
advantage. Stage 3 uses the critic at test time to rank candidates, execute
only the top ones, and feed critiques back when all candidates fail.}
\label{fig:overview}
\end{figure}

Figure~\ref{fig:overview} summarises the training and inference pipeline. The coder
$\pi_c(y\mid x)$ is a standard autoregressive policy. The critic
$\pi_v(c,\hat{s}\mid x,y)$ first emits a short critique in natural
language~$c$, followed by a special \texttt{<verdict>} token whose
probability mass on \texttt{pass}/\texttt{fail} is read as the calibrated
score~$\hat{s}$. Placing the critique before the verdict provides a mild
form of reasoning regularisation: in our experiments it improves
calibration relative to a direct regression head while preserving
interpretability.

\subsection{Calibrated advantage from the critic}
\label{sec:method:advantage}

The core of ExecuCritic is a modification of the GRPO advantage that
uses the critic's signal only after checking calibration. For a group
of $K$ rollouts $\{y_i\}$ with execution rewards $R_i$ and critic scores
$\hat{s}_i$, we define the standardised execution advantage
$A_i=(R_i-\mu_R)/(\sigma_R+\epsilon)$ and the standardised critic
advantage $S_i=(\hat{s}_i-\mu_{\hat{s}})/(\sigma_{\hat{s}}+\epsilon)$.
The resulting advantage is
\begin{equation}
\tilde{A}_i \;=\; A_i \;+\; \alpha\,\rho_K\bigl(R,\hat{s}\bigr)\,S_i,
\label{eq:cact-adv}
\end{equation}
where $\rho_K(R,\hat{s})\in[-1,1]$ is the rank correlation within the
group between executor rewards and critic scores, and $\alpha>0$ is a
fixed shaping weight. The factor $\rho_K$ is the key design choice: when
the critic is well calibrated on a particular prompt ($\rho_K\!\to\!1$),
the critic term reinforces the executor signal and spreads it across
rollouts; when the critic is miscalibrated or adversarial
($\rho_K\!\le\!0$), the term collapses or changes direction, preventing
unreliable scores from being treated as trustworthy rewards. This
behaviour stabilises joint training before the critic has fully
converged.

\subsection{Critic objective}
\label{sec:method:critic}

The critic is updated on the same rollouts. Let $v_i\in\{0,1\}$ be the
verdict token target derived from~$R_i$, and let $z_i$ denote a compact
failure descriptor obtained from the executor trace when the program
fails (syntax error, assertion mismatch, timeout, import/API error or
runtime exception). We train the critic with a loss that ties its verdict,
relative ordering and textual diagnosis to the executor feedback:
\begin{align}
\mathcal{L}_v =
& -\mathbb{E}_i\bigl[\log\pi_v(v_i\mid x,y_i)\bigr] \nonumber\\
& + \beta\,\mathbb{E}_{(i^+,i^-)}\!\Bigl[\max\!\bigl(0,\,m-(\hat{s}_{i^+}-\hat{s}_{i^-})\bigr)\Bigr] \nonumber\\
& + \gamma\,\mathbb{E}_{i:R_i=0}\bigl[-\log\pi_v(z_i\mid x,y_i)\bigr].
\label{eq:critic-loss}
\end{align}
The first term is the likelihood of the executor verdict. The hinge term
is used only for rollout groups that contain at least one passing program
$i^+$ and one failing program $i^-$; it asks the critic to score the
passing program at least $m$ higher. The last term is applied to failed
rollouts and trains the critique to match the failure type extracted from
the executor trace. In practice, the hinge term avoids a nearly constant
verdict score, and the diagnosis term makes the feedback useful when the
coder resamples during inference.

\subsection{Coder objective}
\label{sec:method:coder}

The coder is optimised with a clipped PPO/GRPO surrogate using the
advantage in Eq.~\eqref{eq:cact-adv}:
\begin{equation}
\mathcal{L}_c\!=\!-\mathbb{E}_i\!\!\sum_t\!\!\min\!\Bigl[\,r_{i,t}\,\tilde{A}_i,\;
\mathrm{clip}(r_{i,t},1{-}\varepsilon,1{+}\varepsilon)\,\tilde{A}_i\Bigr]
+\lambda\,\mathrm{KL}\bigl(\pi_c\,\Vert\,\pi_{\text{ref}}\bigr),
\label{eq:coder-loss}
\end{equation}
where $r_{i,t}=\pi_c(y_{i,t}\mid x,y_{i,<t})/\pi_{c,\text{old}}(y_{i,t}\mid x,y_{i,<t})$
is the importance ratio and~$\pi_{\text{ref}}$ is the SFT checkpoint used
as a reference. The KL term, set with $\lambda$ small but nonzero,
prevents the coder from drifting into degenerate styles that the
critic happens to overrate.

\subsection{Training pipeline}
\label{sec:method:pipeline}

ExecuCritic is implemented in three stages.

\smallskip\noindent\textit{Stage 1: SFT initialisation.}
We fine tune the coder with supervision on a mixture of OSS-Instruct
~\cite{wei2023magicoder} and Evol-Code~\cite{luo2024wizardcoder} until
it reliably produces outputs that compile and can be tested. The critic
is initialised on triplets $(x,y,v,z)$, where~$v$ is the executor verdict
and $z$ is either a distilled critique from a stronger teacher
(\emph{e.g.}\ DeepSeek-V3) or a normalised failure descriptor from
the executor trace. This stage requires only a few thousand examples and
avoids learning diagnostic language entirely from scratch.

\smallskip\noindent\textit{Stage 2: Joint RLVR.} On each prompt we draw
$K{=}8$ coder rollouts, score them with both the executor and the
current critic, and apply Eqs.~\eqref{eq:cact-adv}--\eqref{eq:coder-loss}.
The two LoRA adapters are updated sequentially within each optimiser
step: first the critic, then the coder using calibrated advantages
computed from the same rollout-level critic scores, while sharing the
rollouts and the frozen backbone. Algorithm~\ref{alg:cact} gives the full
procedure.

\smallskip\noindent\textit{Stage 3: Refinement at inference.}
At inference, the coder produces $N$~candidates; the critic ranks them,
and only the top~$k$ are executed in the sandbox. If none passes, the
critic's textual critique on the failure with the highest score is fed
back to the coder as additional context, and a new round of $N$~candidates
is sampled. We cap the loop at $T$~rounds. Because critic scoring requires
a single forward pass whereas sandbox execution can take seconds, ranking
with the critic substantially reduces running time relative to executing
all sampled candidates.

\begin{algorithm}[t]
\small
\caption{ExecuCritic joint training (one outer iteration)}
\label{alg:cact}
\begin{algorithmic}[1]
\Require dataset $\mathcal{D}$, executor $E$, coder $\pi_c$, critic $\pi_v$, group size $K$
\For{minibatch of prompts $\{x^{(b)}\}\subset\mathcal{D}$}
    \For{each prompt $x$}
        \State Sample $K$ rollouts $\{y_i\}\sim\pi_c(\cdot\mid x)$
        \State Compute executor rewards and failure descriptors $(R_i,z_i)\!\gets\!E(x,y_i)$
        \State Compute critic scores, critiques and verdict logits with $\pi_v(\cdot\mid x,y_i)$
        \State Compute $A_i$, $S_i$ and the rank correlation $\rho_K(R,\hat{s})$
        \State Form calibrated advantage $\tilde{A}_i \gets A_i+\alpha\,\rho_K\,S_i$
    \EndFor
    \State Update $\pi_v$ with $\mathcal{L}_v$ from Eq.~\eqref{eq:critic-loss}
    \State Update $\pi_c$ with $\mathcal{L}_c$ from Eq.~\eqref{eq:coder-loss} using $\tilde{A}_i$
\EndFor
\end{algorithmic}
\end{algorithm}

\section{Experiments}
\label{sec:exp}

\subsection{Setup}
\label{sec:exp:setup}

\subsubsection{Backbones.} We use \textbf{Qwen3-8B}~\cite{yang2025qwen3} as
the headline backbone and \textbf{DeepSeek-Coder-V2-Lite-Instruct} (16B MoE,
${\sim}2.4$B active)~\cite{zhu2024deepseek} to check that the result carries
across architectures, plus the smaller Qwen3-4B in scaling ablations. All
training uses rank-$32$ LoRA adapters on attention and MLP projections with
the backbone frozen, in bf16 with gradient checkpointing.

\subsubsection{Benchmarks.} We evaluate on eight benchmark views covering
increasingly difficult forms of code intelligence. \textbf{HumanEval+}
and \textbf{MBPP+}~\cite{liu2023your} test short function synthesis
with strengthened unit tests. \textbf{LiveCodeBench} v5~\cite{jain2025livecodebench}
measures competitive programming with contamination controls.
\textbf{BigCodeBench-Hard}~\cite{zhuo2025bigcodebench} stresses library
use, long instructions and reasoning over multiple calls. \textbf{APPS}
~\cite{hendrycks2021measuring} and \textbf{CodeContests}
~\cite{li2022competition} cover longer contest-style programs with more
ambiguous intermediate credit. \textbf{SWE-bench Lite}~\cite{jimenez2024swe}
evaluates issue resolution at repository scale on 300 real GitHub tasks.
\textbf{Multi-SWE-bench} evaluates repository repair across Python and
Java subsets. We further include a \textbf{SWE-bench Verified ranking}
view, where the critic ranks a small set of candidate patches before
execution, and a view for selecting which generated tests should be
trusted as auxiliary verifiers. Function synthesis benchmarks report
unbiased pass@1 with greedy decoding; repository benchmarks report
resolved issue rate; ranking views report NDCG@5 and AUC.

\subsubsection{Training data.} The initial SFT stage uses a
mixture of 75\,k examples from OSS-Instruct~\cite{wei2023magicoder} and
Evol-Code-Alpaca, filtered to keep examples that compile in the sandbox.
RLVR prompts are drawn from the train splits of the Code Contests
corpus~\cite{li2022competition}, the SWE-Gym training set~\cite{pan2024training},
a 30\,k prompt subset of APPS~\cite{hendrycks2021measuring}, and a 12\,k
subset of repository edits whose tests run in Docker within 120 seconds.
We deduplicate against all evaluation benchmarks at the prompt level using
MinHash with Jaccard threshold $0.7$, and remove near duplicate function
signatures by exact AST normalisation. Critic failure descriptors are
normalised into seven categories: syntax, import/API, type, assertion,
timeout, runtime exception and unknown.

\subsubsection{Baselines.} We compare ExecuCritic with: \textbf{(B1)}~the
base backbone without task training; \textbf{(B2)}~a coder trained only
with SFT; \textbf{(B3)}~RLVR with a single coder, using vanilla GRPO and
executor rewards to reproduce the SWE-RL/AceCoder recipe;
\textbf{(B4)}~a prompted system with a frozen coder and frozen critic in
the style of AgentCoder; \textbf{(B5)}~Self-Refine~\cite{madaan2023self}
with the same coder; \textbf{(B6)}~trained critic reranking, with the coder
left at the SFT checkpoint; \textbf{(B7)}~execution of all sampled
candidates across refinement rounds; and \textbf{(B8)}~outcome reward model
shaping, which replaces the textual diagnosis with a scalar reward model.
Trained baselines (B2, B3, B6, B8) share our backbone, prompts and token
budget; prompted ones (B1, B4, B5, B7) reuse the backbone untrained. B3
uses $K{=}16$ rollouts to match or exceed our training-time sandbox calls.

\subsubsection{Hyperparameters.} Unless stated otherwise we use
$K{=}8$ rollouts per prompt, $\alpha{=}0.5$, $\beta{=}0.5$,
$\gamma{=}0.2$, margin $m{=}0.2$, PPO clipping
$\varepsilon{=}0.2$, KL coefficient $\lambda{=}0.01$, learning rate
$5\!\times\!10^{-6}$ for both adapters, batch size~$128$, and a total
of $400$ RLVR steps. At inference we use $N{=}8$, $k{=}3$, $T{=}3$ for
the loop that refines candidates after failed execution. For repository
tasks, each candidate is allowed at most 120 seconds of test execution
and 16k input tokens.

\subsection{Main results}
\label{sec:exp:main}

\begin{table}[t]
\centering
\caption{Function synthesis and programming benchmarks (pass@1, \%), with a
shared backbone and training budget. Best in bold, second best underlined.
HE+, MBPP+, LCB and BCB-H are HumanEval+, MBPP+, LiveCodeBench and
BigCodeBench-Hard.}
\label{tab:func}
\setlength{\tabcolsep}{3pt}
\small
\begin{tabular}{lccccccc}
\toprule
\textbf{Method (Qwen3-8B)} & HE+ & MBPP+ & LCB & BCB-H & APPS & CC & Avg.\\
\midrule
B1 Base direct & 71.3 & 65.8 & 24.7 & 28.1 & 19.4 & 18.8 & 38.0\\
B2 SFT only & 74.6 & 68.4 & 26.2 & 30.0 & 21.1 & 20.5 & 40.1\\
B4 Prompted reviewer system & 75.9 & 69.5 & 27.0 & 31.2 & 22.0 & 21.4 & 41.2\\
B5 Self-Refine & 76.2 & 70.0 & 27.6 & 31.6 & 22.4 & 21.9 & 41.6\\
B6 Trained critic rerank & 77.4 & 71.2 & 28.3 & 32.4 & 23.0 & 22.5 & 42.5\\
B7 Execute all samples & 79.8 & 73.0 & 30.4 & 34.0 & 24.7 & 24.0 & 44.3\\
B3 RLVR with one coder & \underline{80.5} & \underline{73.6} & \underline{30.9} & \underline{34.5} & \underline{25.2} & \underline{24.6} & \underline{44.9}\\
\textbf{Ours: ExecuCritic} & \textbf{84.2} & \textbf{77.0} & \textbf{34.1} & \textbf{37.8} & \textbf{28.8} & \textbf{28.0} & \textbf{48.3}\\
\midrule
\textbf{Method (DS-Coder-V2-Lite)} & HE+ & MBPP+ & LCB & BCB-H & APPS & CC & Avg.\\
\midrule
B3 RLVR with one coder & \underline{82.1} & \underline{74.8} & \underline{31.5} & \underline{35.8} & \underline{26.0} & \underline{25.4} & \underline{45.9}\\
\textbf{Ours: ExecuCritic} & \textbf{85.6} & \textbf{77.9} & \textbf{34.7} & \textbf{38.6} & \textbf{29.5} & \textbf{28.7} & \textbf{49.2}\\
\bottomrule
\end{tabular}
\end{table}

\begin{table}[t]
\centering
\caption{Repository tasks and critic evaluation. Repair columns report
resolve rate; Rank and Test-AUC report NDCG@5 and AUC; Exec. is sandbox
calls per solved instance.}\label{tab:repo}
\setlength{\tabcolsep}{2.2pt}
\small
\begin{tabular}{@{}lcccccc@{}}
\toprule
\textbf{Method (Qwen3-8B)} & SWE-Lite & MSB-Py & MSB-Java & Rank & Test-AUC & Exec.\\
\midrule
B2 SFT only & 9.7 & 8.2 & 5.9 & 42.5 & 57.1 & 17.8\\
B4 Prompted reviewer system & 11.0 & 9.1 & 6.4 & 48.8 & 60.4 & 16.9\\
B5 Self-Refine & 11.7 & 9.6 & 6.8 & 50.2 & 61.0 & 16.1\\
B6 Trained critic rerank & 12.4 & 10.0 & 7.1 & 59.6 & 67.5 & 13.4\\
B7 Execute all samples & 14.0 & 11.3 & 7.9 & -- & -- & 24.0\\
B3 RLVR with one coder & \underline{14.6} & \underline{11.7} & \underline{8.3} & 55.8 & 64.2 & 18.6\\
\textbf{Ours: ExecuCritic} & \textbf{18.3} & \textbf{14.4} & \textbf{10.6} & \textbf{66.9} & \textbf{72.8} & \textbf{10.7}\\
\bottomrule
\end{tabular}
\end{table}

Tables~\ref{tab:func} and~\ref{tab:repo} report the main results. On
Qwen3-8B, ExecuCritic improves over RLVR with a single coder by $+3.4$
average points across the six function and programming benchmarks and by
$+3.7$ on SWE-bench Lite; on DeepSeek-Coder-V2-Lite the margin remains
$+3.3$, so the gain is not an artefact of a weak baseline. The critic also
raises patch ranking NDCG@5 by $+11.1$ and cuts sandbox executions per
solved task from $18.6$ to $10.7$.

The gains are largest on \emph{LiveCodeBench}, \emph{APPS},
\emph{CodeContests} and \emph{SWE-bench Lite}, where generations are
longer and credit assignment is noisier. Prompted reviewer and
Self-Refine baselines remain below RLVR with a single coder, indicating
that role decomposition alone is insufficient; the critic must be
calibrated and trained with the coder.

\subsection{Statistical robustness}
\label{sec:exp:robust}

\begin{table}[t]
\centering
\caption{Robustness of the Qwen3-8B headline comparison: mean $\pm$ standard
deviation over three RLVR seeds, with a paired bootstrap 95\% confidence
interval for the gain over RLVR with a single coder.}
\label{tab:robust}
\setlength{\tabcolsep}{4pt}
\small
\begin{tabular}{lcccc}
\toprule
\textbf{Evaluation} & \textbf{B3 RLVR} & \textbf{ExecuCritic} & \textbf{Gain} & \textbf{95\% CI}\\
\midrule
HumanEval+ pass@1 & $80.5{\pm}0.8$ & $84.2{\pm}0.6$ & $+3.7$ & $[+2.4,+5.0]$\\
MBPP+ pass@1 & $73.6{\pm}0.7$ & $77.0{\pm}0.5$ & $+3.4$ & $[+2.1,+4.6]$\\
LiveCodeBench pass@1 & $30.9{\pm}0.6$ & $34.1{\pm}0.7$ & $+3.2$ & $[+1.8,+4.5]$\\
BigCodeBench-Hard pass@1 & $34.5{\pm}0.7$ & $37.8{\pm}0.6$ & $+3.3$ & $[+2.0,+4.7]$\\
SWE-bench Lite resolve & $14.6{\pm}0.8$ & $18.3{\pm}0.9$ & $+3.7$ & $[+1.9,+5.5]$\\
SWE-bench ranking NDCG@5 & $55.8{\pm}1.0$ & $66.9{\pm}1.2$ & $+11.1$ & $[+8.4,+13.8]$\\
\bottomrule
\end{tabular}
\end{table}

Table~\ref{tab:robust} checks whether the gains fall within normal RLVR
variability. Across three seeds, all paired confidence intervals exclude
zero, and ExecuCritic shows slightly lower variance on short programming
benchmarks. The widest interval occurs on SWE-bench Lite because only
300 repository issues are evaluated, but its lower bound remains above
$+1.9$ points, making a seed artefact unlikely.

\subsection{Ablation study}
\label{sec:exp:ablation}

\begin{table}[t]
\centering
\caption{Ablation on Qwen3-8B: average pass@1 over
HE+/MBPP+/LCB/BCB-H/APPS/CC, and SWE-bench Lite resolve rate. Bold marks
the default setting; $K{=}16$ doubles the training sandbox budget.}
\label{tab:ablation}
\setlength{\tabcolsep}{4.5pt}
\small
\begin{tabular}{lcc}
\toprule
\textbf{Variant} & Avg. pass@1 & SWE-Lite\\
\midrule
Full ExecuCritic (default) & \textbf{48.3} & \textbf{18.3}\\
\quad $\alpha{=}0$ (no critic shaping) & 45.2 & 15.0\\
\quad no calibration gate ($\rho_K\!=\!1$) & 46.0 & 15.8\\
\quad $\beta{=}0$ (no pass/fail margin) & 46.3 & 16.1\\
\quad $\gamma{=}0$ (no critique consistency) & 46.7 & 16.4\\
\quad reward model with scalar score & 46.5 & 16.0\\
\quad no shared backbone & 46.9 & 16.7\\
\quad staged critic then coder & 46.2 & 15.9\\
\quad $K{=}4$ rollouts & 46.1 & 15.7\\
\quad $K{=}16$ rollouts & 48.5 & 18.5\\
\quad $T{=}1$ inference refinement & 46.9 & 16.8\\
\bottomrule
\end{tabular}
\end{table}

Table~\ref{tab:ablation} shows that no single component explains the whole
gain. Removing critic shaping nearly reverts to RLVR, while a constant
trust weight ($\rho_K\!=\!1$) hurts when early critic scores are wrong.
Dropping the pass/fail margin weakens discrimination, and dropping critique
consistency weakens inference-time refinement; the scalar reward model
behaves similarly, so textual diagnosis matters when tied to executor
failure types. Sharing the backbone helps but is not decisive, and staged
training loses more, supporting joint updates on one rollout distribution.
Raising $K$ to 16 gains little while doubling sandbox cost, so we use
$K{=}8$, $T{=}3$.

\subsection{Sample efficiency and reward density analysis}
\label{sec:exp:eff}

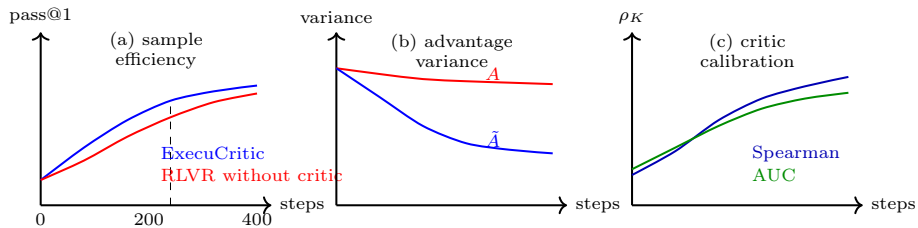
\begin{figure}[t]
\centering
\resizebox{\textwidth}{!}{%
\begin{tikzpicture}[font=\scriptsize]
\begin{scope}[xshift=0cm]
\draw[->, thick] (0,0) -- (3.2,0) node[right] {steps};
\draw[->, thick] (0,0) -- (0,2.4) node[above] {pass@1};
\draw[thick, blue] plot[smooth] coordinates {(0,0.35) (0.6,0.8) (1.2,1.18) (1.8,1.45) (2.4,1.58) (3.0,1.66)};
\draw[thick, red] plot[smooth] coordinates {(0,0.35) (0.6,0.62) (1.2,0.95) (1.8,1.22) (2.4,1.43) (3.0,1.55)};
\draw[dashed] (1.8,0) -- (1.8,1.45);
\node[align=center] at (1.6,2.15) {(a) sample\\efficiency};
\node[blue,anchor=west] at (1.55,0.72) {ExecuCritic};
\node[red,anchor=west] at (1.55,0.42) {RLVR without critic};
\node[below] at (0,0) {0};
\node[below] at (1.5,0) {200};
\node[below] at (3.0,0) {400};
\end{scope}

\begin{scope}[xshift=4.1cm]
\draw[->, thick] (0,0) -- (3.2,0) node[right] {steps};
\draw[->, thick] (0,0) -- (0,2.4) node[above] {variance};
\draw[thick, red] plot[smooth] coordinates {(0,1.9) (0.6,1.82) (1.2,1.75) (1.8,1.72) (2.4,1.70) (3.0,1.68)};
\draw[thick, blue] plot[smooth] coordinates {(0,1.9) (0.6,1.50) (1.2,1.10) (1.8,0.86) (2.4,0.77) (3.0,0.72)};
\node[align=center] at (1.6,2.15) {(b) advantage\\variance};
\node[blue,anchor=west] at (1.95,0.92) {$\tilde{A}$};
\node[red,anchor=west] at (1.95,1.82) {$A$};
\end{scope}

\begin{scope}[xshift=8.2cm]
\draw[->, thick] (0,0) -- (3.2,0) node[right] {steps};
\draw[->, thick] (0,0) -- (0,2.4) node[above] {$\rho_K$};
\draw[thick, blue!70!black] plot[smooth] coordinates {(0,0.42) (0.6,0.75) (1.2,1.18) (1.8,1.48) (2.4,1.65) (3.0,1.78)};
\draw[thick, green!55!black] plot[smooth] coordinates {(0,0.50) (0.6,0.80) (1.2,1.10) (1.8,1.34) (2.4,1.48) (3.0,1.56)};
\node[align=center] at (1.6,2.15) {(c) critic\\calibration};
\node[blue!70!black,anchor=west] at (1.55,0.72) {Spearman};
\node[green!55!black,anchor=west] at (1.55,0.42) {AUC};
\end{scope}
\end{tikzpicture}}
\caption{Sample efficiency and reward density. \textbf{(a)} Pass@1 vs.~GRPO
steps; ExecuCritic reaches the final RLVR accuracy after about 240 steps.
\textbf{(b)} Advantage variance within rollout groups. \textbf{(c)} Critic
calibration (Spearman correlation and pass/fail AUC) during joint training.}
\label{fig:eff}
\end{figure}

Figure~\ref{fig:eff} reports three diagnostics. ExecuCritic reaches the
final accuracy of RLVR with a single coder after roughly $60\%$ of the
policy-gradient steps. Its advantage estimate also becomes less noisy
once~$\rho_K$ exceeds ${\sim}0.4$, giving a more informative gradient.
Meanwhile, the rank correlation between critic scores and executor
outcomes rises from $0.18$ after SFT to $0.71$ at convergence, suggesting
that the critic learns execution correctness rather than coder style.

\subsection{Critic reliability analysis}
\label{sec:exp:reliability}

\begin{table}[t]
\centering
\caption{Reliability of critic variants on validation rollouts. ECE and
Brier measure calibration (lower is better), AUC and Spearman measure
pass/fail discrimination, and Type-F1 measures agreement between the
diagnosed failure type and the executor descriptor.}
\label{tab:reliability}
\setlength{\tabcolsep}{3.2pt}
\small
\begin{tabular}{lccccc}
\toprule
\textbf{Critic variant} & \textbf{ECE}$\downarrow$ & \textbf{Brier}$\downarrow$ & \textbf{AUC}$\uparrow$ & \textbf{Spearman}$\uparrow$ & \textbf{Type-F1}$\uparrow$\\
\midrule
Frozen prompt critic & 0.213 & 0.246 & 60.4 & 0.31 & 38.2\\
Trained critic reranker & 0.124 & 0.184 & 67.5 & 0.52 & 53.0\\
Reward model with scalar score & 0.108 & 0.171 & 68.9 & 0.54 & --\\
\textbf{ExecuCritic critic} & \textbf{0.061} & \textbf{0.137} & \textbf{72.8} & \textbf{0.71} & \textbf{64.6}\\
\bottomrule
\end{tabular}
\end{table}

Table~\ref{tab:reliability} makes the calibration claim more explicit.
We compute ECE on validation rollouts from HumanEval+, LiveCodeBench,
APPS and SWE-bench Lite, and Type-F1 only on failed candidates with a
normalised executor descriptor. A frozen prompt critic is fluent but
often over-predicts pass outcomes on plausible wrong programs. Reranking
training improves discrimination, but joint learning further improves
calibration and diagnostic faithfulness. The scalar reward model obtains
reasonable scores but lacks failure-type diagnoses, explaining its weaker
refinement performance in Table~\ref{tab:ablation}.

\subsection{Execution budget and scaling}
\label{sec:exp:budget}

\begin{table}[t]
\centering
\caption{Accuracy versus cost under fixed inference sandbox budgets on
Qwen3-8B: average pass rate over HE+/MBPP+/LCB/BCB-H/APPS/CC given the
number of executions in the column header. This budget is distinct from
the training rollout group size~$K$.}
\label{tab:pareto}
\setlength{\tabcolsep}{4.5pt}
\small
\begin{tabular}{lccccc}
\toprule
\textbf{Method} & \textbf{2} & \textbf{4} & \textbf{8} & \textbf{16} & \textbf{24}\\
\midrule
Prompted reviewers & 39.8 & 40.7 & 41.2 & 41.6 & 41.8\\
Self-Refine & 40.1 & 41.0 & 41.6 & 42.0 & 42.2\\
Trained critic rerank & 41.5 & 42.2 & 42.5 & 43.0 & 43.2\\
Best of $N$ execution & 40.9 & 42.6 & 44.3 & 45.0 & 45.4\\
RLVR without critic & 43.2 & 44.1 & 44.9 & 44.9 & 45.0\\
\textbf{ExecuCritic} & \textbf{45.8} & \textbf{47.2} & \textbf{48.3} & \textbf{48.5} & \textbf{48.7}\\
\bottomrule
\end{tabular}
\end{table}

Table~\ref{tab:pareto} fixes the maximum sandbox executions at inference.
ExecuCritic dominates the Pareto frontier: with two executions per prompt
it exceeds RLVR with a single coder using eight executions, and with
eight executions it surpasses exhaustive execution of 24 candidates. This
matters most for repository repair, where each sandbox call may require
dependency installation and full test execution.

\begin{table}[t]
\centering
\caption{Execution budget and backbone scaling. Pass@1 is averaged over
HE+/MBPP+/LCB/BCB-H/APPS/CC. Exec. denotes average sandbox calls per
prompt at inference.}
\label{tab:budget}
\setlength{\tabcolsep}{3.2pt}
\small
\begin{tabular}{lcccc}
\toprule
\textbf{Backbone / method} & Params & Exec. & Avg. pass@1 & SWE-Lite\\
\midrule
Qwen3-4B + RLVR without critic & 4B & 16.0 & 39.7 & 10.8\\
Qwen3-4B + ExecuCritic & 4B & 7.2 & \textbf{43.1} & \textbf{13.9}\\
Qwen3-8B + RLVR without critic & 8B & 16.0 & 44.9 & 14.6\\
Qwen3-8B + ExecuCritic & 8B & 7.5 & \textbf{48.3} & \textbf{18.3}\\
DS-Coder-V2-Lite + RLVR without critic & 16B MoE & 16.0 & 45.9 & 15.4\\
DS-Coder-V2-Lite + ExecuCritic & 16B MoE & 7.9 & \textbf{49.2} & \textbf{19.1}\\
\bottomrule
\end{tabular}
\end{table}

Table~\ref{tab:budget} shows that ExecuCritic is not simply buying
accuracy with more sampling. Across all three backbones, it uses less
than half as many sandbox calls as executing all $N$ candidates while
improving pass@1 and repository resolve rate. The relative gain is
slightly larger for Qwen3-4B, where the coder makes more detectable local
mistakes.

\subsection{Case study by bug type}
\label{sec:exp:case}

\begin{table}[t]
\centering
\caption{Failure types on a stratified 250 problem subset of LiveCodeBench
and APPS. Cells give the fraction of problems failing with that bug type
(\%, lower is better); rows do not sum to 100, since a solved problem is
assigned no failure type.}
\label{tab:bugtypes}
\setlength{\tabcolsep}{4pt}
\small
\begin{tabular}{lccccc}
\toprule
\textbf{Failure type} & B2 & B4 & B5 & B3 & \textbf{Ours}\\
\midrule
Boundary / index error & 18.4 & 16.9 & 15.6 & 14.0 & \textbf{10.2}\\
Import or wrong API & 14.1 & 11.5 & 10.8 & 9.7 & \textbf{6.4}\\
Type / numerical precision & 9.8 & 9.2 & 8.7 & 7.6 & \textbf{5.5}\\
Assertion mismatch & 20.6 & 19.4 & 18.9 & 16.8 & \textbf{13.9}\\
Runtime exception & 8.5 & 7.9 & 7.5 & 6.7 & \textbf{4.8}\\
Timeout / complexity & 7.2 & 6.9 & 6.6 & 6.1 & \textbf{4.7}\\
Deep logic / algorithmic & 22.7 & 21.0 & 20.6 & 18.3 & \textbf{15.1}\\
\bottomrule
\end{tabular}
\end{table}

To examine \emph{which} errors the critic helps fix, we label each error
on a stratified 250-problem subset of LiveCodeBench and APPS, using the
taxonomy of \cite{jain2025livecodebench,jiang2026foe} and our executor
descriptors.
Table~\ref{tab:bugtypes} reports the remaining error rate by failure
type. RLVR with a single coder removes many easy bugs but struggles with
deep logic mistakes and rarer runtime failures. ExecuCritic improves all
seven categories, with the largest relative gains on \emph{import/API}
($-34\%$), \emph{runtime exception} ($-28\%$) and
\emph{type/numerical} ($-28\%$) errors, which the critic also flags most
reliably in its textual outputs.

\subsection{Discussion and limitations}
\label{sec:exp:disc}

ExecuCritic inherits two limitations of RLVR. First, it needs an executable
test oracle per training prompt; refactoring or documentation would require
synthetic verifiers or a learned test generator. Second, although
$\rho_K$ in Eq.~\eqref{eq:cact-adv} limits badly calibrated critic rewards,
weak or out-of-distribution backbones may need a longer SFT phase. We saw
no reward hacking, plausibly because the critic is anchored to executor
verdicts, but adversarial initial critics deserve study, since agents can
be steered by inputs crafted to look
benign~\cite{qian2026penny,lou2026helpers}.

\section{Conclusion}
\label{sec:conclusion}

We presented ExecuCritic, a joint reinforcement learning framework in which
a coder and an execution-grounded critic share a backbone and are updated
together under verifiable rewards. A calibrated advantage makes sparse RLVR
feedback denser, while critic ranking and failure diagnosis stabilise
training and support efficient inference. Across eight benchmark views and
two open backbones it beats strong RLVR, prompted reviewer, self-refinement
and scalar reward model baselines using fewer policy-gradient steps and
sandbox executions, and our analyses indicate that learning a calibrated
critic alongside the coder is a useful inductive bias. Future work includes
tasks without test oracles, sparse expert backbones, process reward models,
progressively trained smaller backbones~\cite{liu2026reasonact}, domains
with weaker verifiers~\cite{kang2026quanteval,kang2025jurisctc,zheng2026cite},
and keeping the critic calibrated as the coder
drifts~\cite{feng2026forever,kang2026orderprobe}.

%
%
\bibliographystyle{splncs04}
\bibliography{references}

@article{shao2024deepseekmath,
  title={Deepseekmath: Pushing the limits of mathematical reasoning in open language models},
  author={Shao, Zhihong and Wang, Peiyi and others},
  journal={arXiv:2402.03300},
  year={2024}
}

@article{guo2025deepseek,
  title={Deepseek-r1: Incentivizing reasoning capability in llms via reinforcement learning},
  author={Guo, Daya and Yang, Dejian and others},
  journal={arXiv:2501.12948},
  year={2025}
}

@article{wei2026swe,
  title={Swe-rl: Advancing llm reasoning via reinforcement learning on open software evolution},
  author={Wei, Yuxiang and Duchenne, Olivier and others},
  journal={NeurIPS},
  volume={38},
  pages={78500--78525},
  year={2026}
}

@inproceedings{zeng2025acecoder,
  title={Acecoder: Acing coder rl via automated test-case synthesis},
  author={Zeng, Huaye and Jiang, Dongfu and others},
  booktitle={Proc. ACL},
  year={2025}
}

@article{liu2025code,
  title={Code-r1: Reproducing r1 for code with reliable rewards},
  author={Liu, Jiawei and Zhang, Lingming},
  journal={arXiv:2503.18470},
  year={2025}
}

@article{liu2026rstar,
  title={rStar-Coder: Scaling Competitive Code Reasoning with a Large-Scale Verified Dataset},
  author={Liu, Yifei and Zhang, Li Lyna and others},
  journal={NeurIPS},
  volume={38},
  pages={58780--58807},
  year={2026}
}

@article{pan2024training,
  title={Training software engineering agents and verifiers with swe-gym},
  author={Pan, Jiayi and Wang, Xingyao and others},
  journal={arXiv:2412.21139},
  year={2024}
}

@article{huang2023agentcoder,
  title={Agentcoder: Multi-agent-based code generation with iterative testing and optimisation},
  author={Huang, Dong and Zhang, Jie M and others},
  journal={arXiv:2312.13010},
  year={2023}
}

@inproceedings{hong2024metagpt,
  title={MetaGPT: Meta programming for a multi-agent collaborative framework},
  author={Hong, Sirui and Zhuge, Mingchen and others},
  booktitle={Proc. ICLR},
  year={2024}
}

@inproceedings{islam2024mapcoder,
  title={Mapcoder: Multi-agent code generation for competitive problem solving},
  author={Islam, Md Ashraful and Ali, Mohammed Eunus and others},
  booktitle={Proc. ACL},
  year={2024}
}

@inproceedings{wu2024autogen,
  title={Autogen: Enabling next-gen LLM applications via multi-agent conversations},
  author={Wu, Qingyun and Bansal, Gagan and others},
  booktitle={Proc. COLM},
  year={2024}
}

@article{madaan2023self,
  title={Self-refine: Iterative refinement with self-feedback},
  author={Madaan, Aman and Tandon, Niket and others},
  journal={NeurIPS},
  volume={36},
  pages={46534--46594},
  year={2023}
}

@article{shinn2023reflexion,
  title={Reflexion: Language agents with verbal reinforcement learning},
  author={Shinn, Noah and Cassano, Federico and others},
  journal={NeurIPS},
  volume={36},
  pages={8634--8652},
  year={2023}
}

@article{chen2022codet,
  title={Codet: Code generation with generated tests},
  author={Chen, Bei and Zhang, Fengji and others},
  journal={arXiv:2207.10397},
  year={2022}
}

@article{mcaleese2024llm,
  title={Llm critics help catch llm bugs},
  author={McAleese, Nat and Pokorny, Rai Michael and others},
  journal={arXiv:2407.00215},
  year={2024}
}

@inproceedings{lightman2024let,
  title={Let's verify step by step},
  author={Lightman, Hunter and Kosaraju, Vineet and others},
  booktitle={Proc. ICLR},
  year={2024}
}

@article{cobbe2021training,
  title={Training verifiers to solve math word problems},
  author={Cobbe, Karl and Kosaraju, Vineet and others},
  journal={arXiv:2110.14168},
  year={2021}
}

@inproceedings{ng1999policy,
  title={Policy invariance under reward transformations: Theory and application to reward shaping},
  author={Ng, Andrew Y and Harada, Daishi and others},
  booktitle={Proc. ICML},
  year={1999},}

@article{yuan2024self,
  title={Self-rewarding language models},
  author={Yuan, Weizhe and Pang, Richard Yuanzhe and others},
  journal={arXiv:2401.10020},
  year={2024}
}

@inproceedings{wang2024math,
  title={Math-shepherd: Verify and reinforce llms step-by-step without human annotations},
  author={Wang, Peiyi and Li, Lei and others},
  booktitle={Proc. ACL},
  year={2024}
}

@article{bai2022constitutional,
  title={Constitutional ai: Harmlessness from ai feedback},
  author={Bai, Yuntao and Kadavath, Saurav and others},
  journal={arXiv:2212.08073},
  year={2022}
}

@article{schulman2017proximal,
  title={Proximal policy optimization algorithms},
  author={Schulman, John and Wolski, Filip and others},
  journal={arXiv:1707.06347},
  year={2017}
}

@article{hu2022lora,
  title={Lora: Low-rank adaptation of large language models.},
  author={Hu, Edward J and Shen, Yelong and others},
  journal={Proc. ICLR},
  year={2022}
}

@article{wei2023magicoder,
  title={Magicoder: Empowering code generation with oss-instruct},
  author={Wei, Yuxiang and Wang, Zhe and others},
  journal={arXiv:2312.02120},
  year={2023}
}

@inproceedings{luo2024wizardcoder,
  title={Wizardcoder: Empowering code large language models with evol-instruct},
  author={Luo, Ziyang and Xu, Can and others},
  booktitle={Proc. ICLR},
  year={2024}
}

@article{yang2025qwen3,
  title={Qwen3 technical report},
  author={Yang, An and Li, Anfeng and others},
  journal={arXiv:2505.09388},
  year={2025}
}

@article{zhu2024deepseek,
  title={Deepseek-coder-v2: Breaking the barrier of closed-source models in code intelligence},
  author={Zhu, Qihao and Guo, Daya and others},
  journal={arXiv:2406.11931},
  year={2024}
}

@article{grattafiori2024llama,
  title={The llama 3 herd of models},
  author={Grattafiori, Aaron and Dubey, Abhimanyu and others},
  journal={arXiv:2407.21783},
  year={2024}
}

@article{liu2023your,
  title={Is your code generated by chatgpt really correct? rigorous evaluation of large language models for code generation},
  author={Liu, Jiawei and Xia, Chunqiu Steven and others},
  journal={NeurIPS},
  volume={36},
  pages={21558--21572},
  year={2023}
}

@inproceedings{jain2025livecodebench,
  title={Livecodebench: Holistic and contamination free evaluation of large language models for code},
  author={Jain, Naman and Gu, Alex and others},
  booktitle={Proc. ICLR},
  year={2025}
}

@inproceedings{zhuo2025bigcodebench,
  title={Bigcodebench: Benchmarking code generation with diverse function calls and complex instructions},
  author={Zhuo, Terry Yue and Vu, Minh Chien and others},
  booktitle={Proc. ICLR},
  year={2025}
}

@inproceedings{jimenez2024swe,
  title={Swe-bench: Can language models resolve real-world github issues?},
  author={Jimenez, Carlos E and Yang, John and others},
  booktitle={Proc. ICLR},
  year={2024}
}

@article{li2022competition,
  title={Competition-level code generation with alphacode},
  author={Li, Yujia and Choi, David and others},
  journal={Science},
  volume={378},
  number={6624},
  pages={1092--1097},
  year={2022},}

@article{hendrycks2021measuring,
  title={Measuring coding challenge competence with apps},
  author={Hendrycks, Dan and Basart, Steven and others},
  journal={arXiv:2105.09938},
  year={2021}
}

@inproceedings{zheng2026gradients,
  title={Gradients Know What Outcomes Don't: Unlocking Reinforcement Learning for LLM Reasoning with Gradient-Aligned Rewards},
  author={Zheng, L. and Su, J. and others},
  booktitle={Proc. EMNLP},
  year={2026}
}

@inproceedings{shi2026spader,
  title={SPADER: Step-wise Peer Advantage with Diversity-Aware Exploration Rewards for Multi-Answer Question Answering},
  author={Shi, Q. and Kang, Z. and others},
  booktitle={Proc. EMNLP},
  year={2026}
}

@inproceedings{kang2026jpo,
  title={JPO: Juris Policy Optimization for Structured Legal Reasoning in Criminal Judgment Prediction},
  author={Kang, Z. and Liu, Y. and others},
  booktitle={Proc. EMNLP},
  year={2026}
}

@inproceedings{wang2026ngrams,
  title={Beyond N-grams: A Hierarchical Reward Learning Framework for Clinically-Aware Medical Report Generation},
  author={Wang, Y. and Gao, S. and others},
  booktitle={Proc. AAAI},
  year={2026}
}

@inproceedings{hong2026conflict,
  title={Conflict-Aware Client Selection for Multi-Server Federated Learning},
  author={Hong, M. and Lin, Z. and others},
  booktitle={Proc. ICASSP},
  year={2026}
}

@inproceedings{jiang2026foe,
  title={FoE: Forest of Errors Makes the First Solution the Best in Large Reasoning Models},
  author={Jiang, K. and Dong, H. and others},
  booktitle={Proc. ACL},
  year={2026}
}

@inproceedings{dong2026neureasoner,
  title={NeuReasoner: Towards Explainable, Controllable, and Unified Reasoning via Mixture-of-Neurons},
  author={Dong, H. and Jiang, K. and others},
  booktitle={Proc. ACL},
  year={2026}
}

@inproceedings{jiang2026agentqmix,
  title={Agent Q-Mix: Selecting the Right Action for LLM Multi-Agent Systems through Reinforcement Learning},
  author={Jiang, E.H. and Li, L. and others},
  booktitle={Proc. COLM},
  year={2026}
}

@inproceedings{meng2026group,
  title={Group Cognition Learning: Making Everything Better Through Governed Two-Stage Agents Collaboration},
  author={Meng, C. and Feng, P. and others},
  booktitle={Proc. ICML},
  year={2026}
}

@inproceedings{kang2026mmlegal,
  title={Multimodal Multi-Agent Empowered Legal Judgment Prediction},
  author={Kang, Z. and Gong, J. and others},
  booktitle={Proc. ICASSP},
  year={2026}
}

@inproceedings{zheng2026scilens,
  title={SciLENS: RL-Driven Autonomous Agents for Scientific Localized Evidence Navigation and Synthesis},
  author={Zheng, L. and Su, J. and others},
  booktitle={Proc. EMNLP},
  year={2026}
}

@article{zhu2026edis,
  title={EDIS: Diagnosing LLM Reasoning via Entropy Dynamics},
  author={Zhu, C. and Wu, S. and others},
  journal={arXiv:2602.01288},
  year={2026}
}

@inproceedings{kang2026orderprobe,
  title={How Order-Sensitive Are LLMs? OrderProbe for Deterministic Structural Reconstruction},
  author={Kang, Z. and He, Y. and others},
  booktitle={Findings of EMNLP},
  year={2026}
}

@inproceedings{feng2026forever,
  title={FOREVER: Forgetting Curve-Inspired Memory Replay for Language Model Continual Learning},
  author={Feng, Y. and Wang, H. and others},
  booktitle={Proc. ACL},
  year={2026}
}

@inproceedings{zheng2026cite,
  title={What Should I Cite? A RAG Benchmark for Academic Citation Prediction},
  author={Zheng, L. and Zhang, J. and others},
  booktitle={Proc. WWW},
  year={2026}
}

@article{kang2026quanteval,
  title={QuantEval: A Benchmark for Financial Quantitative Tasks in Large Language Models},
  author={Kang, Z. and Gong, J. and others},
  journal={arXiv:2601.08689},
  year={2026}
}

@inproceedings{zhao2026masbench,
  title={MAS-Bench: A Unified Benchmark for Shortcut-Augmented Hybrid Mobile GUI Agents},
  author={Zhao, P. and Liu, G. and others},
  booktitle={Proc. ACL},
  year={2026}
}

@inproceedings{qian2026penny,
  title={Penny Wise, Pixel Foolish: Bypassing Price Constraints in Multimodal Agents via Visual Adversarial Perturbations},
  author={Qian, J. and Kang, Z.},
  booktitle={Findings of ACL},
  year={2026}
}

@inproceedings{lou2026helpers,
  title={When Helpers Become Hazards: A Benchmark for Analyzing Multimodal LLM-Powered Safety in Daily Life},
  author={Lou, X. and Xu, J. and others},
  booktitle={Findings of ACL},
  year={2026}
}

@inproceedings{liu2026reasonact,
  title={ReasonAct: Progressive Training for Fine-Grained Video Reasoning in Small Models},
  author={Liu, J. and Kang, Z.},
  booktitle={Proc. AAAI},
  year={2026}
}

@inproceedings{kang2025jurisctc,
  title={JurisCTC: Enhancing Legal Judgment Prediction via Cross-Domain Transfer and Contrastive Learning},
  author={Kang, Z. and Cai, H. and others},
  booktitle={Proc. IJCNN},
  year={2025}
}
\end{document}